\documentclass[11pt,rnote] {aa}

\usepackage{graphicx, subfig}
\usepackage{eqnarray}
\usepackage{multirow}
\usepackage{hhline}
\usepackage{siunitx}
\usepackage{natbib,twoopt, units}
\usepackage{lscape}
\usepackage[breaklinks=true]{hyperref} 
\bibpunct{(}{)}{;}{a}{}{,} 
\newcommandtwoopt{\citeads}[3][][]{\href{http://adsabs.harvard.edu/abs/}
{\citealp[#1][#2]{#3}}} 
\newcommandtwoopt{\citepads}[3][][]{\href{http://adsabs.harvard.edu/abs/#3}
{\citep[#1][#2]{#3}}} 
\newcommandtwoopt{\citetads}[3][][]{\href{http://adsabs.harvard.edu/abs/#3}
{\citet[#1][#2]{#3}}} 
\newcommandtwoopt{\citeyearads}[3][][]
{\href{http://adsabs.harvard.edu/abs/#3}{\citeyear[#1][#2]{#3}}}

\title{Elemental ratios in stars vs planets}

\authorrunning{Thiabaud et al.}
\titlerunning{Elemental ratios in stars vs planets}

\author{Amaury~Thiabaud \inst{1,2}, Ulysse~Marboeuf\inst{1,2}, Yann Alibert \inst{1,2,4}, Ingo Leya\inst{1,2} \& Klaus Mezger \inst{1,3} }
\institute{$^1$Center for Space and Habitability, Universit\"{a}t Bern, CH-3012 Bern, Switzerland.\\   
Email: amaury.thiabaud@csh.unibe.ch\\
$^2$Physikalisches Institut, Universit\"{a}t Bern, CH-3012 Bern, Switzerland\\
$^3$Institut f\"{u}r Geologie, Universit\"{a}t Bern, CH-3012 Bern, Switzerland \\
$^4$On leave from CNRS, Observatoire de Besan\c{c}on,  France }
  
\date{Received ??; accepted ??}

\begin{document}

	\abstract
   	{The chemical composition of planets is an important constraint for planet formation and subsequent differentiation. While theoretical studies try to derive the compositions of planets from planet formation models in order to link the composition and formation process of planets, other studies assume that the elemental ratios in the formed planet and in the host star are the same.} 
   	{Using a chemical model combined with a planet formation model, we aim to link the composition of stars with solar mass and luminosity with the composition of the hosted planets. For this purpose, we study the three most important elemental ratios that control the internal structure of a planet: Fe/Si, Mg/Si, and C/O.}
   	{A set of 18 different observed stellar compositions was used to cover a wide range of these elemental ratios. The Gibbs energy minimization assumption was used to derive the composition of planets, taking stellar abundances as proxies for nebular abundances, and to generate planets in a self-consistent planet formation model. We computed the elemental ratios Fe/Si, Mg/Si and C/O in three types of planets (rocky, icy, and giant planets) formed in different protoplanetary discs, and compared them to stellar abundances.}
   	{We show that the elemental ratios Mg/Si and Fe/Si in planets are essentially identical to those in the star. Some deviations are shown for planets that formed in specific regions of the disc, but the relationship remains valid within the ranges encompassed in our study. The C/O ratio shows only a very weak dependence on the stellar value.}   	
	{}
	
   
   	\maketitle
	
	\section{Introduction}
		The composition of planets is important for an understanding of the planet formation process, and the evolution and structure of planets. While the composition of the atmosphere of planets can be observed through the absorption of the star light, for example, the composition of the solids remains difficult to estimate. There are two main hypotheses used to estimate the rocky composition of exoplanets. In the first hypothesis, it is assumed that the solids formed in equilibrium from the material of the solar nebula \citep{Elser2012, Madhusudhan2012a,Thiabaud2013}, so that equilibrium chemistry can be used as a constraint for modelling. The second hypothesis assumes that the elemental ratios Fe/Si and Mg/Si  have stellar values \citep[e.g.,][]{Valencia2010, Wang2013} and these values are then used to model the planetary structure. This last assumption well reproduces the formation and composition of Mars \citep{Khan2008} and Earth \citep[][PREM]{Javoy2010} but not that of present-day Mercury, which has elemental ratios that are not solar, possibly due to its collisional history \citep{Benz2007} or peculiar formation \citep{Lewis1972, Lewis1988}.
				
		Elemental ratios are important because they govern the distribution and formation of chemical species in the protoplanetary disc. Mg/Si governs the distribution of silicates. If the Mg/Si mass ratio is lower than 0.86, Mg is incorporated in orthopyroxene (MgSiO$_{3}$) and the remaining Si enters other silicate minerals such as feldspars (CaAl$_{2}$Si$_{2}$O$_{8}$, NaAlSi$_{3}$O$_{8}$) or olivine (Mg$_{2}$SiO$_{4}$). For Mg/Si ratios from 0.86 to 1.73, Mg is distributed between pyroxene and olivine. For Mg/Si higher than 1.73, Si is incorporated into olivine, and the remaining Mg enters other minerals, mostly oxides. 
	
		The amount of carbides and silicates formed in planets is controlled by the C/O ratio \citep{Larimer1975}. For mass ratios higher than 0.6,  Si is expected to combine with C and form carbides (SiC), because the system is C-rich. If the C/O ratio is lower than 0.6, Si combines with O to form silicates, which are basic building blocks of rock-forming minerals,for instance SiO.\footnote{Note that the Mg/Si and C/O ranges given here are valid for pressures of ca. \num{1e-4} bars. The pressures in protoplanetary discs are lower at most of the distances to the star.} However, as a result of the chemical properties of C (used to form carbohydrates, CO, CO$_2$, etc.), the gaseous C/O ratio in planets can deviate from the stellar value depending on different parameters (temperature, pressure, oxidation state, etc.) and processes during planet formation, including the initial location of formation of the planetary embryos, the migration path of the planet, and the evolution of the gas phase of a protoplanetary disc \citep{Oberg2011,Ali-dib2014, Madhusudhan2014,Thiabaud2014}. \\
	
		In the present study, we show that assuming the elemental Fe/Si and Mg/Si ratios in planets to be of stellar value agrees with the formation of material by condensation from the stellar nebula. To follow this assumption, we quantified the abundance of refractory compounds incorporated into planets for different compositions of stellar nebulae using equilibrium chemistry. From these computed compositions, the elemental ratios Fe/Si, Mg/Si, and C/O in solids (minerals and ices) were derived for three types of planets: rocky, gas giant, and icy planets. These ratios were then compared to the ratios in the host star given in the literature, to determine whether it is valid to assume that ratios are equal in stars and planets. The goal of this study is not to predict the abundances of elements in the planets, but to provide relationships between the star and the planetary composition for stars of solar mass and luminosity with different compositions. In Section 2, we recall the model used to determine the chemical composition of planets. Section 3 is dedicated to the sample of stars from which the composition of the stellar nebula is derived. Results and discussions are presented in Section 4. Section 5 summarizes the conclusions.

	\section{Planet formation and composition model}
	
		The planet formation model used in this study is that of \cite{Alibert2013} (hereafter A13) and is based on the core accretion model. Planets are formed in a protoplanetary disc by accretion of solid planetesimals until they eventually become massive enough to gravitationally bind some of the nebular gas, surrounding themselves with a tenuous atmosphere.
		
		\subsection{Disc model}
		The model of a protoplanetary disc is structured around three modules \citep{Alibert2005d}, which calculate the structure and the evolution of a non-irradiated \citep[see][and references therein]{Alibert2013} and a fully irradiated \citep{Fouchet2012, Thiabaud2013, Marboeuf2014a}\footnote{The irradiation is included by modifying the temperature boundary condition at the disc surface, following \cite{Hueso2005}.} protoplanetary disc. The disc parameters are computed using a 1+1D model, which first resolves the vertical structure of the disc for each distance to the star to derive a vertically averaged viscosity. The calculation includes the hydrostatic equilibrium, the energy conservation, and the diffusion equation for radiative flux. The radial evolution is then computed as a function of this vertically averaged viscosity, photo-evaporation, and mass accretion rate by solving the diffusion equation.
		
		The initial gas surface density (in g.cm$^{-2}$) is computed following
		
		\begin{eqnarray}
			\label{sigma_ini}	
				\Sigma=\Sigma_{0}\left(\frac{r}{a_{0}}\right)^{-\gamma}e^{(\frac{r}{a_{core}})^{2-\gamma}},
		\end{eqnarray}
		
		\noindent where a$_0$ is equal to 5.2 AU, a$_{\rm core}$ is the characteristic scaling radius (in AU), $\gamma$ is the power index, $r$ the distance to the star (in AU), and $\Sigma_0$=(2-$\gamma$)$\frac{M_{\rm disc}}{2\pi a_{\rm core}^{2-\gamma}a_0^\gamma}$ (in g.cm$^{-2}$). The values for a$_{\rm core}$, $\gamma$, and $\Sigma_0$ are varied for every disc and are derived from the observations of \cite{Andrews2010}. The mass of the gas disc is inferred from the observations of solid discs through a dust-to-gas ratio randomly chosen to lie between 0.008 and 0.1 from a list of $\sim$1000 CORALIE targets (see A13 for more details) to produce 500 different discs, whose characteristics are presented in Table \ref{charadisc}.\\
		
		The parametric surface density profile (Eq. \ref{sigma_ini}) is only used as an initial condition (t=0). This initial profile evolves with time, and the surface density profile at t!= 0 is recomputed at each time step, following
		\begin{eqnarray}
			\label{sigma_evol}
			\frac{\delta \Sigma}{\delta t}=\frac{3}{r}\frac{\delta}{\delta r} \left(r^{\rm 1/2} \frac{\delta}{\delta r}(\nu \Sigma r^{\rm 1/2})\right)+Q_{\rm acc}+Q_{\rm ph},
		\end{eqnarray}
		
		\noindent where $Q_{\rm acc}$ is an additional term related to the accretion by planets and $Q_{\rm ph}$ an additional term related to photo-evaporation.
				
		\subsection{Planet formation}
		The simulations of planet formation start with a protoplanetary disc, composed of gas with planetesimals, and ten planetary embryos of lunar mass. These embryos are uniformly distributed (in log) between 0.04 and 30 AU. The processes of planetesimal formation are not the topic of this contribution, and the model describes processes after their formation. Planetesimals are set to be large enough ($\sim$ 1km) so that their drift can be neglected.
		The model includes several processes: 
			\begin{itemize}
				\item accretion of planetesimals from which grows the solid core of the planet \citep[][]{Alibert2005d,Fortier2013}.
				\item accretion of gas. If the core of the planet reaches the critical core mass, the gas is accreted in a runaway fashion, leading to the formation of gas giants \citep[][]{Alibert2005d}.
				\item planet-disc interactions, which contributes to migration (type I or type II) of planets \citep[][]{Alibert2005d,Mordasini2009,Mordasini2009a}.
				\item planet-planetesimal interactions, by which the planetesimal is excited by the planet \citep[][]{Fortier2013}.
				\item gravitational planet-planet interactions, making it possible to catch a pair of planets (or more) in mean motion resonance, as well as pushing each other. 
			\end{itemize}
		The calculations are stopped once the gas disc has dissipated. The cumulative distribution of disc lifetimes is assumed to decay exponetially with a characteristic time of 2.5 Ma \citep[][A13]{Fortier2013}. The photoevaporation rate is then adjusted so that the protoplanetary disc mass reaches 10$^{-5}$M$_\odot$ at the selected disc lifetime, when calculations are stopped. The longest lifetime of a disc is 10 Ma (see A13 for more details).  
				
			\subsection{Refractory composition}
				The chemical model is similar to that used in \cite{Thiabaud2013,Thiabaud2014,Marboeuf2014,Marboeuf2014a}. However, in contrast to these studies, the assumption that all discs have the same pressure profile (equal to the average pressure profile of all discs) for computing the chemical equilibrium is not used\footnote{The implications of this assumption have been discussed in \cite{Thiabaud2013}.}. The presence and abundances of refractory minerals in planetesimals is ruled by the assumption of equilibrium condensation from the primordial stellar nebula. The software HSC Chemistry (v.7.1) computes this equilibrium using a Gibbs energy minimization routine; the pressure and temperature profiles are provided by the planet formation model for each disc. The composition of species in the disc is computed for the initial profiles of $\Sigma$, T, and P and is assumed not to change.
				As in \cite{Thiabaud2013}, the refractory elements have condensed by the time the system has cooled below 200 K; the amount of refractory species at a lower temperature is thus given by their abundance at 200 K. 
		
				\cite{Thiabaud2013} showed that a model in which  refractory organic compounds are not taken into account better reproduces the observation of the solar system. Consequently, the results shown in the present study do not consider the formation of such species, even though observations on single objects (such as Haley's comet) suggest a high carbon inventory that is due to these species, which could be the result of surface effects. The inclusion and discussions of refractory organic material can be found in \cite{Thiabaud2013}.
		
			\subsection{Volatile composition}
				\subsubsection{Volatile components in ices}
				The chemical composition of eight volatile components (CO, CO$_2$, N$_2$, NH$_3$, H$_2$O, CH$_3$OH, CH$_4$, H$_2$S) in planets is determined by condensation and gas-trapping in water ice (e.g. clathrates) on the surface of refractory grains. The thermodynamic conditions of these processes are determined through the partial pressures of the volatile molecules in the gas phase of the disc and by the temperature and surface density of the disc. If the partial pressure of a molecule is higher than the equilibrium pressure of condensation, then the species condenses and its amount at the distance r to the star is determined by the following equation \citep{Mousis2004}:
				
				\begin{eqnarray}
					m_x(r)= \frac{Y_x}{Y_{H_2O}} \times \frac{\Sigma(T_x,P_x,r)}{\Sigma(T_{H_2O},P_{H_2O},r)},
					\label{clath}
				\end{eqnarray}
				
				where Y$_x$ is the mass ratio of the molecules $x$ relative to H$_2$ and Y$_{H_2O}$ the mass ratio of H$_2$O relative to H$_2$ in the disc;  $\Sigma(T_x,P_x,r)$ and $\Sigma(T_{H_2O},P_{H_2O},r)$ are the surface densities of the disc at a distance $r$ to the star, when the molecule $x$ ($\neq$H$_2$O) is condensed or trapped. \\
				
				In previous studies by \cite{Marboeuf2014,Marboeuf2014a}, two models for which the initial CO/H$_2$O molar ratio is 0.2 or 1 were computed, along with the presence of clathrates. The formation of clathrates in protoplanetary discs is uncertain\footnote{The kinetics for clathrate formation are poorly constrained at low temperatures and pressures, and the kinetics feasibility remains uncertain (see \cite{Marboeuf2015} for more details)}, and the change of the initial CO/H$_2$O does not significantly influence the abundances of refractory elements, including the Fe/Si and Mg/Si ratios. Moreover, the C/O ratio is not significantly changed in planets, as shown in \cite{Thiabaud2014}, if the initial CO/H$_2$O is equal to 0.2 or 1. For these reasons, clathrates are not included, and we only computed the model with an initial CO/H$_2$O = 0.2.
				
				\subsubsection{Volatile components in the gas phase}
				The volatile molecules considered in the solid phase are also considered in the gas phase of the disc. The computation of their abundances is similar to that of the ices and is calculated following:
				
				\begin{eqnarray}
		 			N_{\rm X} = Y_{\rm X} \  .\ N_{\rm H_2},
		 		\end{eqnarray}
				
				where $N_{\rm X}$ is the abundance in the gas of species X, $Y_{\rm X}$ is the ISM abundance of species X relative to H$_2$, and $N_{\rm H_2}$ is the abundance of $H_2$.
				
				In contrast to the solid phase, in which the planetesimals are assumed to be large enough not to drift, the gas composition evolves with time following the provision given by the theory of accretion discs by \cite{Lynden-Bell1974}. The radial velocity of the gas $v_{\rm adv}$ (in cm.s$^{-1}$) due to mass and angular momentum conservation in the presence of a viscosity $\nu$ is calculated using
				\begin{eqnarray} \label{equa_drift}
					v_{\rm adv} = - \frac{3}{\Sigma r^{\rm 1/2}} \frac{\delta}{\delta r} (\nu \Sigma r^{\rm 1/2}).
				\end{eqnarray}
				
				The calculation of these abundances in the gas phase is the same as that of \cite{Thiabaud2014}. In the latter, the importance of planet migration and of the gas phase evolution was demonstrated, so that these effects are also included in this study. The authors also showed that it is possible to enhance the C/O ratio of the gas disc to up to three times the stellar value, depending on the time and location, which makes the formation of specific objects, such as the enstatite chondrite parent bodies possible (see their Fig. 8).
		 
	\section{Sample}
		To quantify the relationship between elemental ratios in stars of solar mass and luminosity and planets, a sample set of observed compositions from 18 stars was selected for this study. This sample set encompasses a wide range of Fe/Si (from 1.34 to 2.81), Mg/Si (from 0.66 to 1.85), and C/O (from 0.12 to 0.75) mass ratios. The central elemental abundances in these stars and their central elemental ratios are given in Table \ref{ab_sample} and Table \ref{ratio_sample}.
				
		 \begin{table}[ht]
			\centering
			\caption{\label{ratio_sample} Elemental ratios of stars in the sample set, in mass and moles.}
			\begin{tabular}{|c|c|c|c|c|c|c|}
				\hline
				Star & \multicolumn{2}{c|}{C/O} & \multicolumn{2}{c|}{Mg/Si} & \multicolumn{2}{c|}{Fe/Si}\\
				\hline
				 & Mass & Mol & Mass & Mol & Mass & Mol  \\
				\hline
				\hline
				Sun & 0.37 & 0.50 & 0.88 & 1.02 & 1.69 & 0.86 \\
				55 Cancri & 0.75 & 1.00 & 1.51 & 1.74  & 1.81 & 0.91\\
				HD 967 & 0.19 & 0.25 &1.17& 1.35 & 1.54 & 0.78\\
				HD 1581 & 0.32 & 0.43 & 0.97 & 1.12 & 2.39 & 1.20\\
				HD 2071 & 0.36 & 0.48 &0.93 &1.07 & 2.56 & 1.29 \\
				HD 4308 & 0.29 & 0.39 &1.14 & 1.32 & 1.77 & 0.89\\
				HD 6348 & 0.38 & 0.50 &1.07 & 1.23 & 2.13 & 1.07\\
				HD 9826 & 0.47 & 0.63 & 1.02 & 1.18 & 1.72 & 0.87\\
				HD 19034 & 0.24 & 0.32 &1.245 & 1.45 & 1.61 & 0.81 \\
				HD 37124 & 0.31 & 0.41 & 1.46 & 1.70 & 1.34 & 0.68 \\
				HD 66428 & 0.45 & 0.60 & 1.85 & 2.14 & 2.39 & 1.20\\
				HD 95128 & 0.41 & 0.54 & 0.86 & 1.00 & 1.73 & 0.87\\
				HD 142022 & 0.45 & 0.60 & 1.40 & 1.62 & 2.57 & 1.29\\
				HD 147513 & 0.12 & 0.16 & 0.66 & 0.76 & 2.18 & 1.10\\
				HD 160691 & 0.56 & 0.74 & 0.88 & 1.02 & 1.69 & 0.85\\
				HD 213240 & 0.34 & 0.45 & 1.27 & 1.48 & 1.99 &1.00\\
				HD 220367 & 0.305 & 0.41 & 1.34 & 1.55 & 2.33 & 1.17\\
				HD 221287 & 0.24 & 0.32 & 0.75 & 0.87 & 2.81 & 1.41 \\
				\hline
			\end{tabular}
		\end{table} 
		
		For each star of the sample set, the composition of the protoplanetary disc was computed as discussed in the previous sections, taking the stellar abundances as proxies for nebular abundances. 
				
	\section{Results and discussions}
			The values for elemental ratios of planets were computed by distinguishing three types of planets: rocky, icy, and giant planets. Each type of planets forms a population, based on its mass and position. Rocky planets were defined to be planets whose mass is lower then 10 M$_\oplus$ and whose semi-major axis is lower than 2 AU. Icy planets were defined as low mass planets below 10 M$_\oplus$ located beyond 2 AU, and giant planets were defined as massive planets with masses greater than 10 M$_\oplus$\footnote{In this definition, Neptune- and Uranus-type planets are not considered as icy planets but as giant planets.}. The mode value\footnote{The mode value of a sample is the value that appears most often in a set of data.} of the composition of each population was computed along with the deviation from this value. The latter was used as the error bar.
			
		\subsection{Mg/Si and Fe/Si ratios}
\begin{figure}
				\includegraphics[width=\columnwidth]{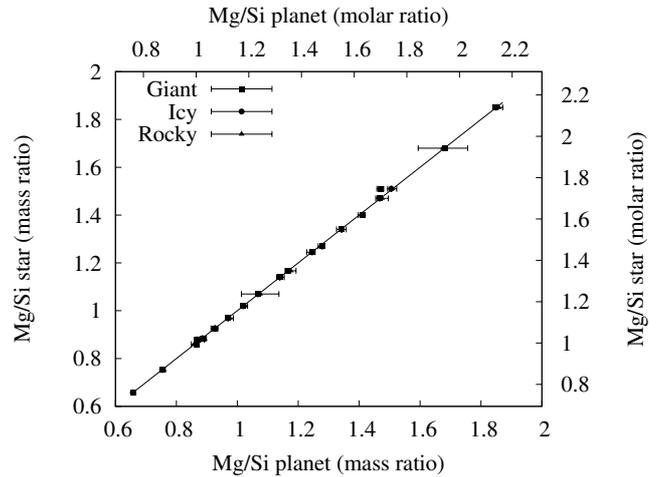}
				\caption{\label{MgSi_nirr} Correlation of Mg/Si in solids accreted by the modelled planets and their respective host star, with irradiation. The line shows the 1:1 relationship. The results for simulations without irradiation are similar.}
			\end{figure}

\begin{figure}
				\includegraphics[width=\columnwidth]{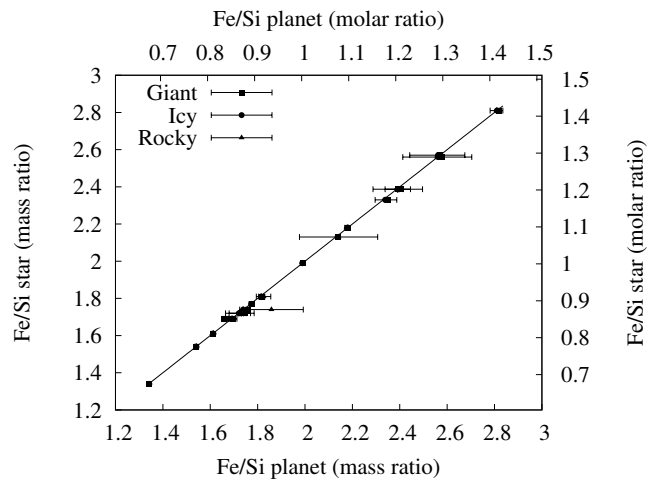}
				\caption{\label{FeSi_nirr} Same as Figure \ref{MgSi_nirr} for Fe/Si.}
			\end{figure}

\begin{figure}
				\includegraphics[width=\columnwidth]{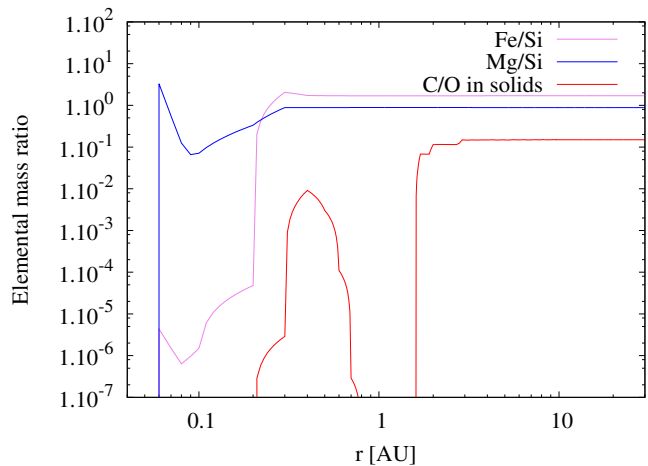}
				\caption{\label{ratio_disc} Elemental mass ratios  Fe/Si, Mg/Si, and C/O in the disc \#1 of A13 ($\Sigma_0$=95.8 g.cm$^{-2}$, a$_{\rm core}$=46 AU, $\gamma$=0.9), with solar composition and without irradiation. At a distance farther than 0.3 AU, the Fe/Si and Mg/Si ratios attain solar values (Fe/Si$_\odot$=1.69; Mg/Si$_\odot$=0.88; C/O$_\odot$=0.37).}
			\end{figure}

				Figure \ref{MgSi_nirr}  provides mode values of Mg/Si and Figures \ref{FeSi_nirr} of Fe/Si in planets formed for each star composition versus the central ratio of the host star as given in Table \ref{ratio_sample} for a case with irradiation. For all three types of planets, and for the two irradiation cases, the element ratios between stars and planets correlate along a 1:1 relationship.				
				This is an expected result for icy and giant planets that mainly formed outside the ice line, a region where all refractory material has condensed.
				Inside the ice line, the elemental ratio in the disc is in some instances different from the stellar value, as shown in Figure \ref{ratio_disc}, since the temperature is too high to condense all Mg- and Fe-bearing species. It is thus possible to form planets with non-stellar ratios, either because of their migration pathway or the location of formation, close to the star, which will result in different element ratios in the planets. However, the formation and migration of most rocky planets inside the ice line mainly occurs in regions where the temperature and the pressure during condensation are such that these ratios will be stellar in the planets. 
							
		\subsection{C/O ratio in solids}
\begin{figure}
				\includegraphics[width=\columnwidth]{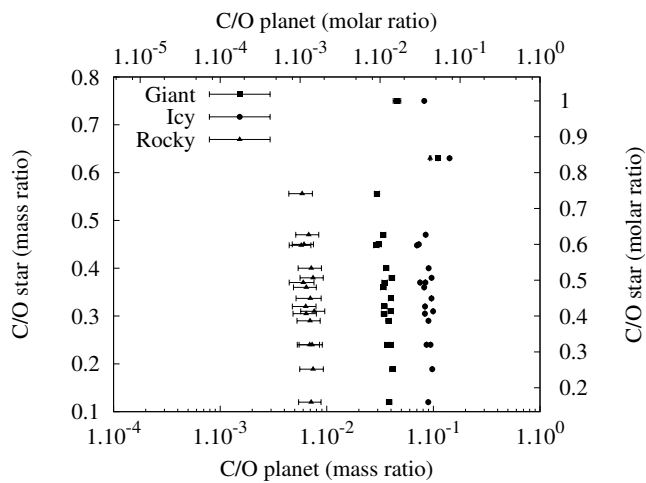}
				\caption{\label{CO_nirr} Correlation of C/O in solids acrreted by the modelled planets and their respective host star, with irradiation. The results for simulations without irradiation are similar with values of roughly one order of magnitude lower.}
			\end{figure}

			Figure \ref{CO_nirr} shows the mode value of C/O in solids (refractory species and ices) for the three populations versus the stellar ratio. The C/O ratio in icy and giant planets appears to be very weakly dependent on the C/O ratio in the star. 
			In contrast to other studies \citep[see, e.g.,][]{Madhusudhan2012a}, an increase of the C/O ratio does not imply a carbon-rich interior in this study, which could be due to the assumption that the composition of the planetesimals does not evolve, as suggested by \cite{Moriarty2014}. \cite{Moriarty2014} also calculated successive condensation steps and their results show that successive condensations favour the formation of C-rich material during later condensations. The largest differences also come from the inclusion of volatile species, which are not taken into account in these other studies, and which governs the amount of C incorporated into planetesimals. Another possibility is that there is not enough carbon to form such planets and that the C/O ratio of the host star needs to be higher for the typical pressures involved in protoplanetary discs.
			
			The difference between the types of planets can be explained by their location of formation and migration pathway. Rocky planets are formed in a region where no ices are formed, thus the abundance of carbon is very low (see Fig. \ref{ratio_disc} -  C is mainly used to form ices). Icy planets form beyond the ice lines, and the ices formed there control the C and O budget \citep[see][]{Marboeuf2014a}. This leads to a higher abundance of carbon and consequently to a higher C/O ratio. Giant planets fall between the two exemples. They are formed in the simulations mainly between the water and CO$_2$ ice lines, thus the C/O ratio in solids is lower than that of the icy planets, but higher than that of the rocky planets. For a model without irradiation, icy and giant planets have similar C/O ratios. In this case, ice lines are closer to each other, which leads to minimal differences between icy and giant planets.
			Consequently, the C/O ratio in solids is mainly constrained by ices for non-terrestrial planets and is limited by the low amount of C in refractory species for terrestrial planets.
			
		\subsection{C/O ratio in gas}
			As discussed in \cite{Thiabaud2014}, the dynamical processes involved in the C/O ratio of the envelopes of planets (migration, gas phase evolution, etc.) enables each type of planets to form gaseous envelopes with a range in their C/O mass ratios, ranging from 0 to 1.2 in the case of solar composition. Because of these processes, which do not depend on the chemical properties of the star, it is impossible to establish a relationship between the stellar composition and the C/O ratio of the envelope of a planet. Indeed, \cite{Thiabaud2014} showed that two planets located at the same distance with the same mass could acquire a different C/O ratio depending on the time of accretion and their formation pathways. It is thus hardly possible to relate the C/O ratio in the envelope of a planet to its host star.
			However, as shown in \cite{Thiabaud2014}, the total C/O ratio (which includes ices and gas) is strongly dominated by ices. Consequently, the results of the total C/O ratio are similar to those presented in the previous section.
			
			\subsection{Effect of T and P}
\begin{figure}
				\includegraphics[width=\columnwidth]{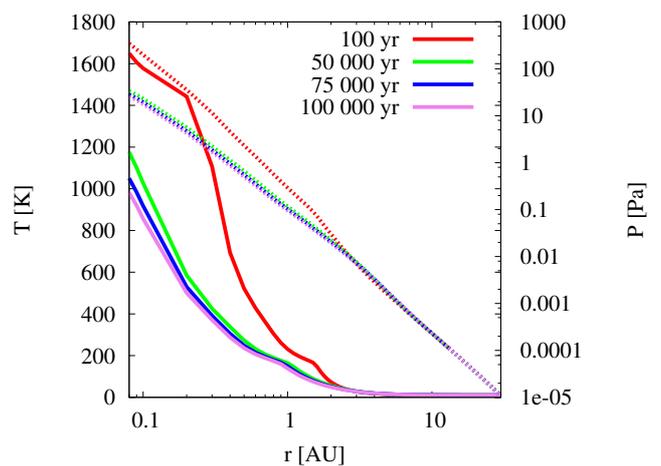}
				\caption{\label{rtp_evol} Time evolution of the mid-plane temperature (solid lines) and pressure (dotted lines) of disc 1 ($\Sigma_0$=95.8 g.cm$^{-2}$, a$_{\rm core}$=46 AU, $\gamma$=0.9) in a case without irradiation. From \cite{Thiabaud2014}.}
			\end{figure}

			Mg/Si and Fe/Si ratios are always stellar  beyond the water iceline since the temperature is low enough to condense all refractory material, which explains the absence of large error bars in Figures \ref{MgSi_nirr} and \ref{FeSi_nirr} for icy and giant planets. However, this is not the case at all locations inside the water iceline. The ratios at these heliocentric distances depend on what has condensed within the disc. Since thermodynamic properties are different among molecules, they condense at different temperatures and pressures, and thus the budget of Fe, Mg, and Si might differ depending on the temperature and pressure profiles of discs, resulting in non-stellar values for Mg/Si and Fe/Si ratios, as shown in Figure \ref{ratio_disc}. Figure \ref{rtp_evol} shows the profiles of temperature and profile for disc 1 of A13. Beyond 0.3 AU, the temperature is low enough ($\sim$ 1000K) so that most of the Mg-, Si- and Fe-bearing species condense, which is not the case closer to the star. Hence ratios diverge from stellar values.
		
		\subsection{In situ formation}
		
		Migration is an important process that affects the chemical composition of a planet, but it can smoothen the compositional differences of bodies formed at different distances to the host star. If planets are formed in situ, then the ratios (and especially the C/O ratio) will be representative of that ratio in a specific location of the disc. The effect of migration is not as strong on the Fe/Si and Mg/Si ratios, since the stellar value is easily obtained (see Fig.3) due to the early condensation of these elements. However, it is stronger for the C/O ratio, because of the volatile character of most C-bearing species, as discussed in Section 4.2.
		
		Considering a giant planet formed at a final position of 5 AU in disc 1 of A13. The water iceline in this disc is located at 1.516 AU, and all the other volatile molecules condense within 4 AU.  In this case, the C/O ratio in the solids at 5 AU will be at its maximum (all C and O has condensed, except for some losses during formation of volatile species), and thus the giant planets, which formed in situ, will have this ratio as well. If the giant planet migrated from 3 AU to 5 AU, into a region where the C/O ratio is lower, a smaller amount of C present in the nebula has condensed compared to the amount of O because less C-bearing volatile species have condensed), then the C/O ratio at the end of the simulation will be lower, resulting in a C/O ratio that would be representative of a mix between what was accreted between 3AU and 5AU. Icy planets and rocky planets migrate less efficiently, resulting in higher (resp. lower) C/O ratios for icy planets (resp. rocky planets), where nearly all the C- and O-bearing species have condensed (resp. nearly none of the C-bearing species have condensed, but some O-bearing species have).

		Since the Mg/Si and Fe/Si ratios have attained a stellar value beyond 0.3 AU in this disc, the \textit{in situ} formation of giant and icy planets will not change, except for hot Jupiters very close to their star. The change will be seen in rocky planets, however. If no migration occurs, the mode value of such ratios for rocky planets will not be stellar.
		
	\section{Summary and conclusions}
		The planet formation model of A13 was used along with the chemical model used in \cite{Marboeuf2014}, \cite{Marboeuf2014a}, and \cite{Thiabaud2013,Thiabaud2014} to derive the composition of planets around 18 different stars with different Fe/Si, Mg/Si, and C/O ratios to deduce their elemental ratios. This study shows that the composition of planets that formed from condensed material from a nebula closely resembles the composition of the host star, as assumed in interior structure models, for solar masses and luminosities.
		
		The Fe/Si and Mg/Si ratios were found to be the same in planets and host star. The C/O ratio in solids was found to be very weakly dependent on the host star ratio for all populations, which contradicts previous studies on the carbon content of terrestrial planets, mainly because these studies did not consider volatile species. The total C/O ratio in planets is governed by ices, and thus the location of planet formation within the evolving planetary disc plays an important role, as does possible migration of these bodies during and after accretion. The C/O acquired in the envelope of a planet can vary greatly between two populations due to physical processes, and it is thus impossible to link this ratio to the composition of the host star. This highlights how complex the origin of volatiles on rocky planets can be and emphasizes that it is necessary to study the origin of volatile components in our solar system by examining meteorites more close.
						
		\paragraph{\textbf{Acknowledgments}} This work was supported by the European Research Council under grant 239605, the Swiss National Science Foundation, and the Center for Space and Habitability of the University of Bern. This work has been in part carried out within the frame of the National Centre for Competence in Research PlanetS supported by the Swiss National Science Foundation.
		
	\bibliographystyle{aa}
	\bibliography{biblio} 

	\appendix
	\section{Appendices}

	\begin{table*}[ht]
			\centering
			\caption{\label{ab_sample} Elemental abundances (in mole) in stars selected for this sample study. The abundances of all elements in all stars are reported relative to an H abundance of \num{1e12} mole. When no values were found for an element, the solar elemental abundances were taken for the calculations. \\
			References: 1. \cite{Lodders2003} 2. \cite{Gilli2006} 3. \cite{Beirao2005} 4. \cite{Ecuvillon2004} 5. \cite{Ecuvillon2006} 6. \cite{DelgadoMena2010} 7. \cite{Neves2009} 8. \cite{Sousa2008} 9. \cite{Gonzalez2007} 10. \cite{Kang2011} 11. \cite{Caffau2005} 12. \cite{Takeda2005} 13. \cite{Morel2004}}
			\begin{tabular}{|c|c|c|c|c|c|c|c|c|}
				\hline
				Star & C & O & Mg & Ni & Si & S & Ti & Fe \\
				\hline
				\hline
				Sun &  \num{2.88e8} &  \num{5.75e8} &  \num{4.17e7} &  \num{1.95e6} &  \num{4.07e7} &  \num{1.82e7} &  \num{1.00e5} &  \num{3.47e7} \\
				55 Cancri & \num{7.40e8} & \num{7.40e8} & \num{1.20e8} & \num{3.60e6} & \num{6.90e7} & \num{2.10e7} & \num{2.20e5} & \num{6.30e7} \\
				HD 967 & \num{1.45e8} & \num{5.75e8} & \num{1.70e7} & \num{3.98e5} & \num{1.26e7} & - & \num{4.17e4} & \num{9.77e6}\\
				HD 1581 & \num{2.29e8} & \num{5.37e8} & \num{2.88e7} & \num{1.12e6} & \num{2.57e7} & - & \num{7.41e4} & \num{3.09e7} \\
				HD 2071 & \num{2.75e8} & \num{5.75e8} & \num{3.16e7} & \num{1.38e6} & \num{2.95e7} & - & \num{8.71e4} & \num{3.80e7}\\
				HD 4308 & \num{2.63e8} & \num{6.76e8} & \num{3.16e7} & \num{8.91e5} & \num{2.40e7} & - & \num{7.59e4} & \num{2.14e7}\\
				HD 6348 & \num{1.91e8} & \num{3.80e8} & \num{1.48e7} & \num{4.57e5} & \num{1.20e7} & - & \num{4.68e4} & \num{1.29e7} \\
				HD 9826 & \num{4.79e8} & \num{7.59e8} & \num{6.38e7} & \num{2.37e6} & \num{5.42e7} & \num{1.74e7} & \num{1.35e5} & \num{4.69e7} \\
				HD 19034 & \num{2.24e8} & \num{6.92e8} & \num{2.75e7} & \num{6.31e5} & \num{1.91e7} & - & \num{6.61e4} & \num{1.55e7} \\
				HD 37124 & \num{2.46e8} & \num{5.96e8} & \num{3.24e7} & \num{7.41e5} & \num{1.91e7} & \num{5.25e6} & \num{7.00e4} & \num{1.29e7} \\
				HD 66428 & \num{6.31e8} & \num{1.05e9} & \num{1.48e8} & \num{3.72e6} & \num{6.92e7} & - & \num{1.78e5} & \num{8.32e7} \\
				HD 95128 & \num{3.84e8} & \num{7.10e8} & \num{4.27e7} & \num{1.86e6} & \num{4.27e7} & \num{1.45e7} & \num{1.14e5} & \num{3.72e7} \\ 
				HD 142022 & \num{4.90e8} & \num{8.20e8} & \num{9.12e7} & \num{2.82e6} & \num{5.62e7} & - & \num{1.70e5} & \num{7.24e7} \\
				HD 147513 & \num{2.40e8} & \num{1.51e9} & \num{3.47e7} & \num{2.19e6} & \num{4.57e7} & - & \num{1.62e5} & \num{5.01e7} \\
				HD 160691 & \num{5.25e8} & \num{7.08e8} & \num{8.51e7} & \num{4.37e6} & \num{8.32e7} & \num{2.57e7} & \num{2.07e5} & \num{7.08e7} \\
				HD 213240 & \num{5.40e8} & \num{1.20e9} & \num{6.50e7} & \num{2.40e6} & \num{4.40e7} & \num{1.30e7} & \num{1.30e5} & \num{4.40e7} \\
				HD 220367 & \num{2.40e8} & \num{5.89e8} & \num{3.80e7} & \num{1.05e6} & \num{2.45e7} & - & \num{7.08e4} & \num{2.88e7} \\
				HD 221287 & \num{3.09e8} & \num{9.77e8} & \num{3.16e7} & \num{1.70e6} & \num{3.63e7} & - & \num{1.12e5} & \num{5.13e7} \\
				\hline
				\hline
				Star  & N & Na & Al & P & Ca & Cr  & \multicolumn{2}{|c|} {Reference} \\	
				\hline
				\hline
				Sun & \num{7.94e7} &  \num{2.34e6} &  \num{3.47e6} &  \num{3.47e5} &  \num{2.57e6} &  \num{5.25e5} & \multicolumn{2}{|c|} {1} \\
				55 Cancri & \num{1.80e8} & \num{3.90e6} & \num{8.70e6} & \num{8.50e5}& \num{2.80e6} & \num{7.80e5} & \multicolumn{2}{|c|} {2, 3, 4, 5} \\
				HD 967  & - & \num{5.50e5} & \num{1.23e6} & - & \num{7.94e5} & \num{1.20e5} & \multicolumn{2}{|c|} {6, 7, 8} \\
				HD 1581 & - & \num{1.38e6} & \num{1.95e6} & - & \num{1.70e6} & \num{3.31e5} & \multicolumn{2}{|c|} {6, 7, 8} \\
				HD 2071  & - & \num{1.66e6} & \num{2.45e6} & - & \num{1.95e6} & \num{3.63e5} & \multicolumn{2}{|c|} {6, 7, 8} \\
				HD 4308  & -  & \num{1.17e6} & \num{2.45e6} & - & \num{1.48e6} & \num{2.14e5} & \multicolumn{2}{|c|} {6, 7, 8} \\
				HD 6348 & - & \num{5.37e5} & \num{1.17e6} & - & \num{7.94e5} & \num{1.48e5} & \multicolumn{2}{|c|} {6, 7, 8} \\
				HD 9826 & \num{8.51e7} & \num{3.94e6} & - & - & \num{3.10e6} & - & \multicolumn{2}{|c|} {9, 11, 12} \\
				HD 19034 & - & \num{9.12e5} & \num{2.04e6} & - & \num{1.20e6} & \num{1.78e5} & \multicolumn{2}{|c|} {6, 7, 8} \\
				HD 37124 & - & \num{9.33e5} & \num{2.29e6} & - & \num{1.15e6} & \num{1.70e5} & \multicolumn{2}{|c|} {9, 10, 11} \\
				HD 66428 & - & \num{4.90e6} & \num{6.31e6} & - & \num{3.72e6} & \num{8.32e5} & \multicolumn{2}{|c|} {6, 7, 8} \\
				HD 95128 & \num{1.05e8} & \num{2.40e6} & \num{3.47e6} & - & \num{2.24e6} & \num{4.90e5} & \multicolumn{2}{|c|} {9, 10, 11, 12} \\
				HD 142022 & - & \num{2.82e6} & \num{5.13e6} & - & \num{3.39e6} & \num{6.76e5} & \multicolumn{2}{|c|} {6, 7, 8} \\
				HD 147513 & - & \num{1.95e6} & \num{3.31e6} & - & \num{3.31e6} & \num{4.90e5} & \multicolumn{2}{|c|} {6, 7, 8} \\
				HD 160691 & - & \num{5.50e6} & \num{7.59e6} & - & \num{4.47e6} & \num{1.01e6} & \multicolumn{2}{|c|} {2, 9, 11, 13}  \\
				HD 213240 & \num{1.40e7} & \num{3.60e6} & \num{4.70e6} & \num{4.60e5} & \num{2.50e6} & \num{5.20e5} & \multicolumn{2}{|c|} {2, 3, 4, 5} \\
				HD 220367 & - & \num{1.29e6} & \num{2.40e6} & - & \num{1.55e6} & \num{2.88e5} & \multicolumn{2}{|c|} {6, 7, 8} \\
				HD 221287 & - & \num{1.86e6} & \num{2.40e6} & - & \num{2.63e6} & \num{5.01e5} & \multicolumn{2}{|c|} {6, 7, 8} \\
				\hline
			\end{tabular}
		\end{table*}
                  	 \begin{table}[ht]
				\centering
				\caption{\label{charadisc} Characteristics of disc models from A13 assuming a gas-to-dust ratio of 100.}
				\begin{tabular}{|c|c|c|c|}
					\hline
					Disc & M$_{disc}$ (M$_{\odot})$ & a$_{core}$ (AU) & $\gamma$ \\
					\hline
					\hline
					1 & 0.029 & 46 & 0.9 \\
					2 & 0.117 & 127 & 0.9 \\
					3 & 0.143 & 198 & 0.7 \\
					4 & 0.028 & 126 & 0.4 \\
					5 & 0.136 & 80 & 0.9 \\
					6 & 0.077 & 153 & 1.0 \\
					7 & 0.029 & 33 & 0.8 \\
					8 & 0.004 & 20 & 0.8 \\
					9 & 0.012 & 26 & 1.0 \\
					10 & 0.007 & 26 & 1.1 \\
					11 & 0.007 & 38 & 1.1 \\
					12 & 0.011 & 14 & 0.8 \\
					\hline
				\end{tabular}
			\end{table} 

\end{document}